# Novel Reconfigurable Logic Gates Using Spin Metal-Oxide-Semiconductor Field-Effect Transistors

Satoshi Sugahara[1, 2*], Tomohiro Matsuno[1], and Masaaki Tanaka[1,2**]

[1] *Department of Electronic Engineering, The University of Tokyo, 7-3-1 Hongo, Bunkyo-ku, Tokyo, 113-8656, Japan*
[2] *PRESTO, Japan Science and Technology Agency, 4-1-8 Honcho, Kawaguchi 332-0012, Japan*

We propose and numerically simulate novel reconfigurable logic gates employing spin metal-oxide-semiconductor field-effect transistors (spin MOSFETs). The output characteristics of the spin MOSFETs depend on the relative magnetization configuration of the ferromagnetic contacts for the source and drain, that is, high current-drive capability in the parallel magnetization and low current-drive capability in the antiparallel magnetization [S. Sugahara and M. Tanaka: Appl. Phys. Lett. **84** (2004) 2307]. A reconfigurable NAND/NOR logic gate can be realized by using a spin MOSFET as a driver or an active load of a complimentary MOS (CMOS) inverter with a neuron MOS input stage. Its logic function can be switched by changing the relative magnetization configuration of the ferromagnetic source and drain of the spin MOSFET. A reconfigurable logic gate for all symmetric Boolean functions can be configured using only five CMOS inverters including four spin MOSFETs. The operation of these reconfigurable logic gates was confirmed by numerical simulations using a simple device model for the spin MOSFETs.

[*]E-mail address: sugahara@cryst.t.u-tokyo.ac.jp
[**]E-mail address: masaaki@ee.t.u-tokyo.ac.jp

# 1  Introduction

The area of spintronics (or spin electronics) in which uses not only charge transport of electrons but also the spin degree of freedom of electrons is used has generated much interest in recent years. One of the most attractive new directions in such spintronics research is the use of the spin degree of freedom in *active* semiconductor devices and integrated circuits, while most of the studies on spintronics have been concentrated on novel magnetic materials and related *passive* devices so far. Since the present electronics system depends entirely on Si integrated circuits, Si-based active spin devices, which are expected to have good compatibility with present Si technology, are extremely important. Very recently, we proposed a new spin metal-oxide-semiconductor field-effect transistor (spin MOSFET)[1] consisting of a MOS structure and half-metallic-ferromagnet (HMF) [2-7] contacts for the source and drain, as shown in Fig. 1(a). When the magnetization configuration of the HMF source and drain is parallel (antiparallel), highly spin-polarized carriers injected from the HMF source to the channel are transported into (blocked by) the HMF drain. Thus, the spin MOSFET shows high (low) current-drive capability in parallel (antiparallel) magnetization, that is, the transconductance $g_m$ is high ($g_m^P$) for parallel magnetization and $g_m$ is low ($g_m^{AP}$) for antiparallel magnetization. Our theoretical calculation revealed that the spin MOSFET not only exhibits magnetization-dependent output characteristics, but also satisfies other requirements for spintronic integrated circuit applications, such as large amplification capability, small power-delay product, and low off-current. Furthermore, its tunable $g_m^P$ and $g_m^{AP}$ lead to various applications of the spin MOSFET[8]. Nonvolatile memory and reconfigurable logic will be important applications of the spin MOSFET. The $g_m^P$ and $g_m^{AP}$ values of the spin MOSFET can be designed by adjusting the device parameters of the HMF source and drain. While a large $g_m^P$ and a negligibly small $g_m^{AP}$ are favorable for nonvolatile memory applications, a large $g_m^P$ and a moderate $g_m^{AP}$ are required for reconfigurable logic applications, as will be shown in this paper.

In this paper, we propose novel spintronic reconfigurable logic gates employing spin MOSFETs. It is shown that a reconfigurable NAND/NOR logic gate can be realized by using a spin MOSFET as a driver or an active load of a complimentary MOS (CMOS) inverter with a neuron MOS (νMOS)[8, 9] input stage. A reconfigurable logic gate for all symmetric Boolean functions (AND, OR, XOR, NAND, NOR, XNOR, all-"0" and all-"1") can also be configured by using only five CMOS inverters including four spin MOSFETs. The operation of these reconfigurable logic gates was analyzed with the numerical circuit simulator HSPICE using a simple device model for spin MOSFETs.

## 2  Device Model for Spin MOSFETs

A simple device model was applied to the spin MOSFET for performing logic simulations. Since $g_m$ ($=\partial I_D/\partial V_{GS}$) of the spin MOSFET shows an approximately linear increase with increasing gate bias $V_{GS}$ when $V_{GS}$ is higher than the threshold voltage $V_t$[1], the output characteristics (the drain current $I_D$) of the spin MOSFET can be approximated as $I_D = \beta(V_{GS}-V_t)^2$, where $\beta$ is the gain coefficient. This means that the ordinary device models for conventional MOSFETs are useful for the spin MOSFET. Assuming that the drain current of the spin MOSFET has the same function formula as that of the gradual channel approximation for MOSFETs, the static transfer characteristics of the logic gates presented in this paper can be expressed by the *gain coefficient ratio* of their load and driver transistors (the transfer characteristics of a conventional CMOS inverter are described in the same manner). Thus, the absolute drain current values of the spin MOSFETs and conventional MOSFETs are not needed in the static operation analysis of the presented logic gates. Instead, the gain coefficient ratios of the load and driver transistors are used in the following simulations.

To reproduce the magnetization-configuration-dependent output characteristics of the spin MOSFET, large and small gain coefficients are introduced into the device model of a single spin MOSFET, that is, the spin MOSFET in parallel (antiparallel) magnetization is represented by a MOSFET with a large (small) gain coefficient $\beta^P$ ($\beta^{AP}$). Since the gain coefficient of MOSFETs is proportional to the device dimension ratio $W / L$ (where $W$ is the channel width and $L$ the channel length), $\beta^P$ and $\beta^{AP}$ for a single spin MOSFET are separately realized by the appropriate choice of $W / L$ values for parallel and antiparallel magnetizations. Note that although the gate capacitance of this spin MOSFET model is changed according to the $W / L$ values for parallel and antiparallel magnetizations, this variation of the gate capacitance can be neglected when using a νMOS input stage, as described later.

    The numerical simulations of the presented logic gates were performed with a commercially available HSPICE simulator that is based on a sophisticated device model to reproduce transistor characteristics and circuit performance with reasonable accuracy. The design rule and the supply voltage $V_{DD}$ used for the simulations were 0.35 μm and 3.3 V, respectively, although the design rule of sub-100 nm and $V_{DD}$ of 1 − 1.5 V would be more realistic for the spin MOSFET[1]. The threshold voltage $V_t$ of the $n$-channel and $p$-channel MOSFETs were 1.2 and 1.3 V, respectively. These values were also used for the spin MOSFET for simplicity. Note that $V_t$ = ~1 V for $V_{DD}$ = 3.3 V can be scaled to $V_t$ = 0.2 − 0.3 V for $V_{DD}$ = 1 − 1.5 V which is a favorable value for sub-100-nm-scale spin MOSFETs[1]. Red and blue curves in Fig. 1(b) show the output characteristics of this spin MOSFET model in parallel and antiparallel magnetizations, respectively, where the $W / L$ values are 8.75 for parallel magnetization and 1.25 for antiparallel magnetization and the resulting $\beta^P / \beta^{AP}$ value is 7.0. The gain coefficient ratio $\beta^P / \beta^{AP}$ of the spin MOSFET is a suitable parameter for controlling the device performance, since $\beta^P / \beta^{AP}$ directly determines the magnetization-configuration-dependent output characteristics of the spin MOSFET as well as the operational margin of the reconfigurable logic gates presented in this paper. Although the operational margin depends on the ratio $\beta^P / \beta^{AP}$, the appropriate range of $\beta^P / \beta^{AP}$ seems to be extremely wide, 3 − 1000, as estimated from our simulated results. Note that the operating speed and power dissipation of the proposed reconfigurable logic gates also depends on the gain

coefficient ratio $\beta^P / \beta^{AP}$, e.g., a large $\beta^P / \beta^{AP}$ ratio results in small power dissipation at the expense of the operating speed.

## 3    Reconfigurable AND/OR Gate

Figure 2 shows a reconfigurable NAND/NOR gate for the output $V_{O1}$, which acts as an AND/OR gate for the output $V_{O2}$.  The NAND/NOR gate can be realized by using a *p*-channel MOSFET as the active load ($Q_1$) and an *n*-channel spin MOSFET as the driver ($Q_2$) of a CMOS inverter with a νMOS input stage having two binary inputs (*A* and *B*).   Here, the gain coefficients of $Q_1$ and $Q_2$ in parallel and antiparallel magnetizations are expressed by $\beta_1$, $\beta_2^P$ and $\beta_2^{AP}$, respectively, and these are set to satisfy $\beta_2^{AP} < \beta_1 < \beta_2^P$, as discussed later.   The function of a CMOS inverter $G_O$ (whose logic threshold voltage is set at $1/2V_{DD}$) in the output stage is also described later.   Note that this logic gate can also be configured with a *p*-channel spin MOSFET as $Q_1$ and an *n*-channel MOSFET as $Q_2$, and with *p*- and *n*-channel spin MOSFETs as $Q_1$ and $Q_2$, respectively.

The νMOS input stage consists of a floating gate coupled capacitively with two input gates.   The floating-gate voltage $V_{FG}$ of the νMOS is given by[9,10]

$$V_{FG} = \frac{C_A A + C_B B}{C_0 + C_A + C_B} = \frac{A+B}{2}, \qquad (1)$$

where $C_0$ denotes a capacitance between the substrate and the floating gate, $C_A$ and $C_B$ represent a coupling capacitance for inputs *A* and *B*, and we assume $C_A = C_B$ and $C_A, C_B \gg C_0$.   The binary input voltages of 0 and $V_{DD}$ for *A* and *B* can be simply expressed by "0" and "1" which are measured in the units of $V_{DD}$ (hereafter, quotation marks are used to denote values measured in the units of $V_{DD}$).   When the input combinations are $A = B =$ "0" and $A = B =$ "1", $V_{FG}$ is "0" and "1", respectively.   When one of the two inputs is "1" ( i.e., $A =$ "0", $B =$ "1", or $A =$ "1", $B =$ "0"), $V_{FG}$ is "1/2".

Red and blue dotted curves in Fig. 2(b) show the transfer characteristics ($V_{O1}/V_{DD}$ vs

$V_{FG}/V_{DD}$) of the reconfigurable logic gate in parallel and antiparallel magnetizations for the spin MOSFET $Q_2$, respectively, where $\beta_2^P/\beta_1 = 2.7$ and $\beta_2^{AP}/\beta_1 = 0.4$.  When the magnetization of the source and drain of $Q_2$ is parallel, the logic threshold voltage $V_T$ is $V_{TL}$, which is lower than "1/2".  By flipping the magnetization of $Q_2$ from the parallel to antiparallel configuration, the logic threshold voltage $V_T$ is changed to $V_{TH}$, which is higher than "1/2" (see Fig. 2(b)).  This can be explained by assuming that the conventional MOSFET model is applicable to the spin MOSFET as follows: The logic threshold voltage $V_T$ of the reconfigurable logic gate is given by

$$V_T = \frac{V_{DD} - |V_{t1}| + V_{t2}\sqrt{\beta_2^\zeta/\beta_1}}{1 + \sqrt{\beta_2^\zeta/\beta_1}}, \qquad (2)$$

where $V_{t1}$ and $V_{t2}$ are the threshold voltages for the drain currents of $Q_1$ and $Q_2$, respectively, and $\zeta$ is denoted P for parallel magnetization and AP for antiparallel magnetization.  When $|V_{t1}| \approx V_{t2}$, $V_T$ is lower than "1/2" for $\beta_2^\zeta/\beta_1 > 1$ and is higher than "1/2" for $\beta_2^\zeta/\beta_1 < 1$.  Thus, when $Q_2$ is in the parallel (antiparallel) magnetization state, the logic threshold voltage is $V_{TL}$ ($V_{TH}$), owing to the above-noted relation, $\beta_2^{AP} < \beta_1 < \beta_2^P$.

The logic operations of the reconfigurable NAND/NOR gate can be obtained using the transfer characteristics shown in Fig. 2(b).  When the spin MOSFET $Q_2$ is in the parallel magnetization state, the logic threshold voltage $V_{TL}$ is lower than "1/2" as shown by the red dotted curve in the figure.  For the input combinations of $A = B =$ "0", {$A =$ "0", $B =$ "1", or $A =$ "1", $B =$ "0"} and $A = B =$ "1", $V_{FG}$ takes "0", "1/2" and "1", respectively, as described above.  These $V_{FG}$ values are transformed to $V_{O1} =$ "1", "0" and "0", respectively, via the transfer characteristics.  Thus, the reconfigurable logic gate shows a NOR function when $Q_2$ is in the parallel magnetization state.  Note that when $V_{FG} =$ "1/2" {$A =$ "0", $B =$ "1", or $A =$ "1", $B =$ "0"}, $V_{O1}$ is slightly higher than "0".  However, this deviation can be included in a logic margin, i.e., using the logic margin $\Delta V$, the "0" level for $V_{O1}$ is given by "0" $\leq V_{O1} \leq$ "0"$+\Delta V$.  This deviation disappears after the inverse amplification from $V_{O1}$ to $V_{O2}$ by the output inverter $G_O$, as shown by the solid red curve in Fig. 2(b), while the logic function is inverted from NOR for the output $V_{O1}$ to OR for $V_{O2}$.

When the spin MOSFET $Q_2$ is in antiparallel magnetization, the logic threshold voltage $V_{TH}$ is higher than $V_{FG}$ = "1/2", as shown by the dotted blue curve in Fig. 2(b). Owing to this transfer characteristics, the input combinations of $A = B$ = "0", {$A$ = "0", $B$ = "1", or $A$ = "1", $B$ = "0"} and $A = B$ = "1" (corresponding to $V_{FG}$ = "0", "1/2" and "1", respectively) are transformed to $V_{O1}$ = "1", "1" and "0". Thus, the reconfigurable logic gate shows a NAND function when $Q_2$ is in the antiparallel magnetization state. Note that when $V_{FG}$ = "1/2" {$A$ = "0", $B$ = "1", or $A$ = "1", $B$ = "0"}, $V_{O1}$ is slightly lower than "1". Since this deviation can be included in the logic margin for the "1" level ("1"$-\Delta V \leq V_{O1} \leq$ "1"), it is also eliminated by the output inverter $G_O$, as shown by the solid blue curve in Fig. 2(b). The logic function for the output $V_{O2}$ is inverted from NAND to AND.

Figure 3 shows another reconfigurable AND/OR gate for the output $V_{O1}$ and a NAND/NOR gate for the output $V_{O2}$, where the capacitively coupled inputs of the νMOS input stage are replaced by CMOS inverters $G_A$ and $G_B$ that are directly connected to the common gate of $Q_1$ and $Q_2$. Since the logic threshold voltages of $G_A$ and $G_B$ are set to $V_T$ = "1/2", the voltage input to the common gate is "1", "1/2" and "0" for the input combinations of $A = B$ = "0", {$A$ = "0", $B$ = "1", or $A$ = "1", $B$ = "0"} and $A = B$ = "1", respectively[11]. Thus, this input stage shows a similar function to that of the νMOS input stage, although the logic functions of the output are inverted by $G_A$ and $G_B$. In practice, a νMOS input stage requires a large area in order to realize a large input capacitance. On the other hand, the input stage shown in Fig. 3 is effective for reducing the occupied area for the reconfigurable logic gate.

## 4  Reconfigurable Logic Gate for All Symmetric Boolean Functions

Figure 4(a) shows a reconfigurable logic gate for all symmetric Boolean functions (AND, OR, XOR, NAND, NOR, XNOR, all-"1", all-"0"), which is composed of only ten transistors (five CMOS inverters) including four spin MOSFETs. A νMOS input stage and a CMOS inverter $G_{12}$ consisting of $p$-channel and $n$-channel spin MOSFETs $Q_1$ and $Q_2$ act as a NAND/NOR gate for the output $V_{O1}$. A CMOS inverter $G_P$ ($G_N$) and a $p$-channel ($n$-channel) spin MOSFET $Q_3$ ($Q_4$) are connected between the floating gate and the output $V_{O1}$ terminal, as shown in the figure. A CMOS inverter $G_O$ at the output stage is used for the inverse amplification of the $V_{O1}$ signal in order to eliminate deviations of $V_{O1}$ from the complete "0" and "1" states. Note that since the $p$-channel and $n$-channel spin MOSFETs $Q_3$ and $Q_4$ can also create a CMOS configuration ($G_{34}$), this logic gate is compatible with CMOS technology. Hereafter, the gain coefficient of the spin MOSFET $Q_i$ ($i$ = 1, 2, 3 and 4) is referred to as $\beta_i^\zeta$, where $\zeta$ represents P for parallel magnetization and AP for antiparallel magnetization. The relationships of $\beta_3^{AP}$, $\beta_4^{AP} < \beta_1^{AP}$, $\beta_2^{AP} < \beta_1^{P}$, $\beta_2^{P} < \beta_3^{P}$, $\beta_4^{P}$ are required, as discussed later. The logic threshold voltage of $G_{12}$ can be controlled by the magnetization configurations of $Q_1$ and $Q_2$. When $Q_1$ ($Q_2$) is parallel magnetization and $Q_2$ ($Q_1$) is antiparallel magnetization, the logic threshold voltage is $V_{TH} >$ "1/2" ($V_{TL} <$ "1/2") owing to $\beta_2^{AP}/\beta_1^{P} < 1$ ($\beta_2^{P}/\beta_1^{AP} > 1$). This relationship is the same qualitatively as the logic gate shown in Fig. 2, i.e., this relationship can be obtained from eq. (2) by replacing $\beta_1$ with $\beta_1^\zeta$. The logic threshold voltages $V_{TP}$ and $V_{TN}$ of the CMOS inverters $G_P$ and $G_N$ are designed to realize the relationships "1/2" $< V_{TP} < V_{TH}$ and $V_{TL} < V_{TN} <$ "1/2", respectively. Owing to the logic threshold voltage of $G_P$ ($G_N$), $Q_3$ ($Q_4$) turns on only when A = B = "1" (A = B = "0"). The logic threshold voltage of the output inverter $G_O$ is set to "1/2", as well as the reconfigurable logic gate shown in Fig. 2. Note that the vMOS input stage can be replaced by the input stage consisting of the inverters $G_A$ and $G_B$ with the common gate, as shown in Fig. 3.

The logic functions of the reconfigurable logic gate shown in Fig. 4(a) can be switched by changing the combination of magnetization configurations of the spin MOSFETs $Q_1$, $Q_2$, $Q_3$ and $Q_4$. Hereafter, the magnetization configurations of $Q_1$, $Q_2$, $Q_3$ and $Q_4$ are specified using the expression of $\{Q_1, Q_2, Q_3, Q_4\} = \{\zeta_1, \zeta_2, \zeta_3, \zeta_4\}$, where $\zeta_i$ ($i$ = 1, 2, 3 and 4) represents the magnetization configuration of $Q_i$ and $\zeta_i$ = P (AP) is for the parallel (antiparallel) magnetization.

Figures 4 (b)-(e) show transfer characteristics for $\{Q_1, Q_2, Q_3, Q_4\} = \{\zeta_1, \zeta_2, AP, AP\}$, $\{\zeta_1, \zeta_2, P, AP\}$, $\{\zeta_1, \zeta_2, AP, P\}$ and $\{\zeta_1, \zeta_2, P, P\}$, respectively, where $\beta_1^P/\beta_1^{AP} = 7.0$, $\beta_2^{AP}/\beta_1^{AP} = 1.2$, $\beta_2^P/\beta_1^{AP} = 8.0$, $\beta_3^{AP}/\beta_1^{AP} = 0.8$, $\beta_3^P/\beta_1^{AP} = 22.5$, $\beta_4^{AP}/\beta_1^{AP} = 0.5$ and $\beta_4^P/\beta_1^{AP} = 14.0$. The dotted and solid curves in the figure are the transfer characteristics for the output $V_{O1}$ and $V_{O2}$, respectively. Although the output $V_{O1}$ shows deviations from the complete "0" and "1" states (i.e., the "0" level for $V_{O1}$ means "0" $\leq V_{O1} \leq$ "0"$+\Delta V$, and the "1" level for $V_{O1}$ "1"$-\Delta V \leq V_{O1} \leq$ "1"), these deviations can be eliminated using the output inverter $G_O$. Since the resulting logic functions for the output $V_{O2}$ are easily obtained from the inverted logic functions for the output $V_{O1}$, the operation of the reconfigurable logic for the output $V_{O1}$ are described below. Note that all symmetric Boolean functions (AND, OR, XOR, NAND, NOR, XNOR, all-"1", all-"0") are also realized for the output $V_{O2}$, as well as $V_{O1}$.

In the following, we will describe in detail how the function of this circuit can be selected from among all symmetric Boolean functions (AND, OR, NAND, NOR, XOR, XNOR, all 1, all 0) by controlling the magnetization (transconductance) of the four spin MOSFETs ($Q_1$, $Q_2$, $Q_3$ and $Q_4$). Table I shows the truth tables of this circuit.

When $\{Q_1, Q_2, Q_3, Q_4\} = \{\zeta_1, \zeta_2, AP, AP\}$, the reconfigurable logic gate can switch NAND/NOR functions, the logic function of which can be selected by the magnetization configuration combination $\{\zeta_1, \zeta_2\}$ (= $\{P, AP\}$ or $\{AP, P\}$). Although $Q_3$ and $Q_4$ turn on for $V_{FG} > V_{TP}$ and $V_{FG} < V_{TN}$, respectively, the influences of $Q_3$ and $Q_4$ can be neglected because of their small gain coefficients (drain current) compared with those of $Q_1$ and $Q_2$, owing to the above noted relationship of $\beta_3^{AP}$, $\beta_4^{AP} < \beta_1^{AP}$, $\beta_2^{AP} < \beta_1^P$, $\beta_2^P$. Thus, this reconfigurable logic gate exhibits qualitatively the same transfer characteristics as the NAND/NOR gate shown in Fig. 2. The blue and red dotted curves in Fig. 4(b) show the transfer characteristics for $\{Q_1, Q_2, Q_3, Q_4\}$ = $\{P, AP, AP, AP\}$ and $\{AP, P, AP, AP\}$, respectively. Since these magnetization configurations realize $\beta_2^{AP}/\beta_1^P < 1$ and $\beta_2^P/\beta_1^{AP} > 1$, the logic threshold voltages of these transfer characteristics become $V_{TH} >$ "1/2 for $\{Q_1, Q_2, Q_3, Q_4\} = \{P, AP, AP, AP\}$ and $V_{TL} <$ "1/2" for $\{Q_1, Q_2, Q_3, Q_4\}$ =

{AP, P, AP, AP}. The transfer characteristics with $V_{TH}$ and $V_{TL}$ realize NAND and NOR functions (Fig. 4(b)), respectively, as well as those in Fig. 2(b).

When $\{Q_1, Q_2, Q_3, Q_4\} = \{\zeta_1, \zeta_2, P, AP\}$, the reconfigurable logic gate can switch all-"1"/XNOR functions. The blue and red dotted curves in Fig. 4(c) show the transfer characteristics for $\{Q_1, Q_2, Q_3, Q_4\} = \{P, AP, P, AP\}$ and $\{AP, P, P, AP\}$, respectively. Since the gain coefficients $\beta_3^P$ and $\beta_4^{AP}$ satisfy the relationship $\beta_4^{AP} < \beta_1^{AP}$, $\beta_2^{AP} < \beta_1^P$, $\beta_2^P < \beta_3^P$, $Q_3$ turns on and affects the output $V_{O1}$ when $V_{FG} > V_{TP}$, but $Q_4$ does not affect the output $V_{O1}$ even when $V_{FG} < V_{TN}$. Thus, when $V_{FG} < V_{TP}$, the transfer characteristics are almost the same as that for $\{Q_1, Q_2, Q_3, Q_4\} = \{\zeta_1, \zeta_2, AP, AP\}$, and when $V_{FG} > V_{TP}$, the output $V_{O1}$ is forced to increase to the "1" level by $Q_3$ due to the large gain coefficient $\beta_3^P$. Owing to these transfer characteristics, the input combinations of $A = B = $ "0", $\{A = $ "0", $B = $ "1", or $A = $ "1", $B = $ "0"$\}$, and $A = B = $ "1" (corresponding to $V_{FG} = $ "0", "1/2" and "1", respectively) are transformed to $V_{O1} = $ "1", "1" and "1" (all-"1" function) for $\{Q_1, Q_2, Q_3, Q_4\} = \{P, AP, P, AP\}$ and to $V_{O1} = $ "1", "0" and "1" (XNOR function) for $\{Q_1, Q_2, Q_3, Q_4\} = \{AP, P, P, AP\}$, respectively.

When $\{Q_1, Q_2, Q_3, Q_4\} = \{\zeta_1, \zeta_2, AP, P\}$, the reconfigurable logic gate can switch XOR/all-"0" functions. The blue and red dotted curves in Fig. 4(d) show the transfer characteristics for $\{Q_1, Q_2, Q_3, Q_4\} = \{P, AP, AP, P\}$ and $\{AP, P, AP, P\}$, respectively. Owing to the relationship $\beta_3^{AP} < \beta_1^{AP}$, $\beta_2^{AP} < \beta_1^P$, $\beta_2^P < \beta_4^P$, $Q_4$ turns on and affects the output $V_{O1}$ when $V_{FG} < V_{TN}$, but $Q_3$ has no influence on the output $V_{O1}$ even when $V_{FG} > V_{TP}$. Thus, when $V_{FG} > V_{TN}$, the transfer characteristics are almost the same as those for $\{Q_1, Q_2, Q_3, Q_4\} = \{\zeta_1, \zeta_2, AP, AP\}$, and when $V_{FG} < V_{TN}$, the output $V_{O1}$ is forced to decrease to the "0" level by $Q_4$ due to the large gain coefficient $\beta_4^P$. The input combinations of $A = B = $ "0", $\{A = $ "0", $B = $ "1", or $A = $ "1", $B = $ "0"$\}$, and $A = B = $ "1" (corresponding to $V_{FG} = $ "0", "1/2" and "1", respectively) are transformed to $V_{O1} = $ "0", "1" and "0" (XOR function) for $\{Q_1, Q_2, Q_3, Q_4\} = \{P, AP, AP, P\}$ and to $V_{O1} = $ "0", "0" and "0" (all-"0" function) for $\{Q_1, Q_2, Q_3, Q_4\} = \{AP, P, AP, P\}$, respectively.

When $\{Q_1, Q_2, Q_3, Q_4\} = \{\zeta_1, \zeta_2, P, P\}$, this logic gate can switch OR/AND functions.

The blue and red dotted curves in Fig. 4(e) show the transfer characteristics for $\{Q_1, Q_2, Q_3, Q_4\}$ = {P, AP, P, P} and {AP, P, P, P}, respectively. $Q_4$ and $Q_3$ turn on when $V_{FG} < V_{TN}$ and $V_{FG} > V_{TP}$, respectively, and there exists the relationship of $\beta_1^{AP}$, $\beta_2^{AP} < \beta_1^{P}$, $\beta_2^{P} < \beta_3^{P}$, $\beta_4^{P}$. Therefore the transfer characteristics are almost the same as those for $\{Q_1, Q_2, Q_3, Q_4\} = \{\zeta_1, \zeta_2, AP, AP\}$ only when $V_{TN} < V_{FG} < V_{TP}$. When $V_{FG} < V_{TN}$ and $V_{FG} > V_{TP}$, the output $V_{O1}$ is shifted to the "0" and "1" levels by $Q_4$ and $Q_3$, respectively. The input combinations of $A = B =$ "0", {$A =$ "0", $B =$ "1", or $A =$ "1", $B =$ "0"}, and $A = B =$ "1" (corresponding to $V_{FG} =$ "0", "1/2" and "1", respectively) are transformed to $V_{O1} =$ "0", "1" and "1" (OR function) for $\{Q_1, Q_2, Q_3, Q_4\} =$ {P, AP, P, P} and to $V_{O1} =$ "0", "0" and "1" (AND function) for $\{Q_1, Q_2, Q_3, Q_4\} =$ {AP, P, P, P}, respectively.

A possible application of the proposed reconfigurable logic gates is in the recently emerging field programmable gate array (FPGA) [12,13]. The most advanced type of FPGA is realized by using look-up tables (LUTs) as its logic blocks. Since SRAM is used for the LUT, logic functions defined in the LUT are reconfigurable but volatile. Thus, nonvolatile memory is required to define the logic functions, which results in the increase of the chip size of FPGA. On the other hand, our reconfigurable logic gates can store the information of logic functions in the form of the magnetization configuration of the spin MOSFETs. Therefore, the logic functions of our reconfigurable logic gates are nonvolatile. A new FPGA architecture without SRAM and without nonvolatile memory for logic blocks can be established by using spin MOSFETs.

This new class of reconfigurable logic gates could be used for reconfigurable computing, which is an emerging new computing paradigm satisfying both flexibility and performance[14]. Since our device (spin MOSFET) and reconfigurable logic gates are predicted to have both high performance and flexibility as well as nonvolatility, we believe that they will find a variety of applications.

## 5  Conclusion

Novel spintronic reconfigurable logic gates employing spin MOSFETs were proposed. All symmetric Boolean functions can be realized using only five CMOS inverters including four spin MOSFETs, and the logic functions can be switched by changing the combination of the magnetization configurations of the spin MOSFETs.  These reconfigurable and nonvolatile logic gates will be fully compatible with the current CMOS technology, and will provide new "spintronic" Si integrated circuit architectures.

Acknowledgements

The authors would like to thank Professor. T. Shibata and Dr. Y. Mita at the University of Tokyo for helpful discussions.  This work was supported by the PRESTO program of Japan Science and Technology Agency, a Giant-in-Aid for Science Research on the Priority Area "Semiconductor Nanospintronics" (14076207), the IT program of RR2002 from Ministry of Education, Culture, Sports, Science and Technology, Toray Science Foundation, and VLSI Design and Education Center (VDEC) at the University of Tokyo in collaboration with Synopsys, Inc.

Figure Captions

Fig. 1 (a) Schematic device structure of a spin MOSFET using half-metallic-ferromagnet contacts for the source and drain. (b) Output characteristics of a spin MOSFET in parallel magnetization (red curve) and antiparallel magnetization (blue curve).

Fig. 2 (a) Circuit configuration of a reconfigurable NAND/NOR gate for the output $V_{O1}$ and a reconfigurable AND/OR gate for the output $V_{O2}$. (b) Calculated transfer characteristics with $Q_2$ in the parallel magnetization state (red curve) and in the antiparallel magnetization state (blue curve), where dotted and solid curves are the transfer characteristics for outputs $V_{O1}$ and $V_{O2}$, respectively.

Fig. 3 Circuit configuration of a reconfigurable AND/OR gate for the output $V_{O1}$ and a reconfigurable NAND/NOR gate for the output $V_{O2}$, using two CMOS inverters as the input stage.

Fig. 4 (a) Circuit configuration of a reconfigurable logic gate for two-input all symmetric Boolean functions. Transfer characteristics ($V_{O1} - V_{FG}$ and $V_{O2} - V_{FG}$ characteristics normalized by $V_{DD}$) for (b) $\{Q_1, Q_2, Q_3, Q_4\} = \{\zeta_1, \zeta_2, AP, AP\}$, (c) $\{\zeta_1, \zeta_2, P, AP\}$, (d) $\{\zeta_1, \zeta_2, AP, P\}$ and (e) $\{\zeta_1, \zeta_2, P, P\}$ are shown. Dotted and solid curves are the transfer characteristics for outputs $V_{O1}$ and $V_{O2}$, respectively, and blue and red curves show the transfer characteristics for $\{\zeta_1, \zeta_2\} = \{P, AP\}$ and $\{AP, P\}$, respectively.

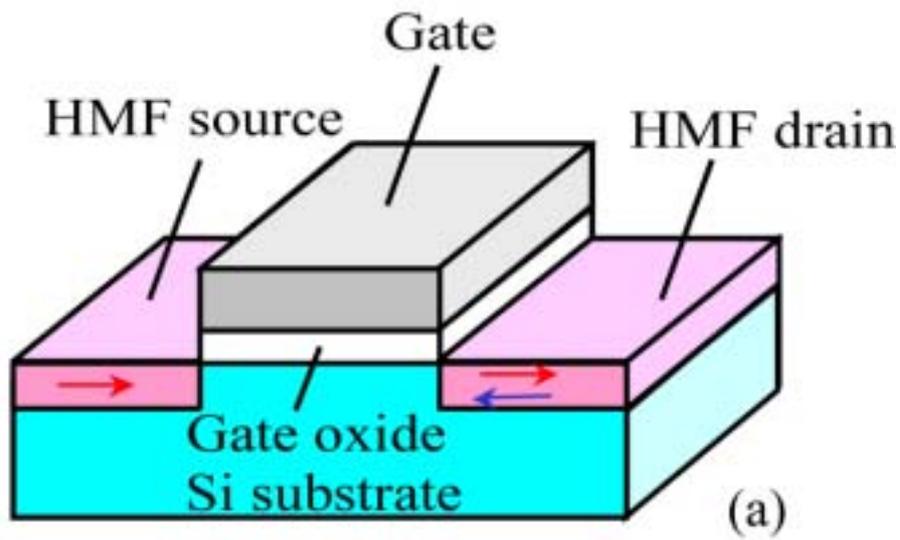

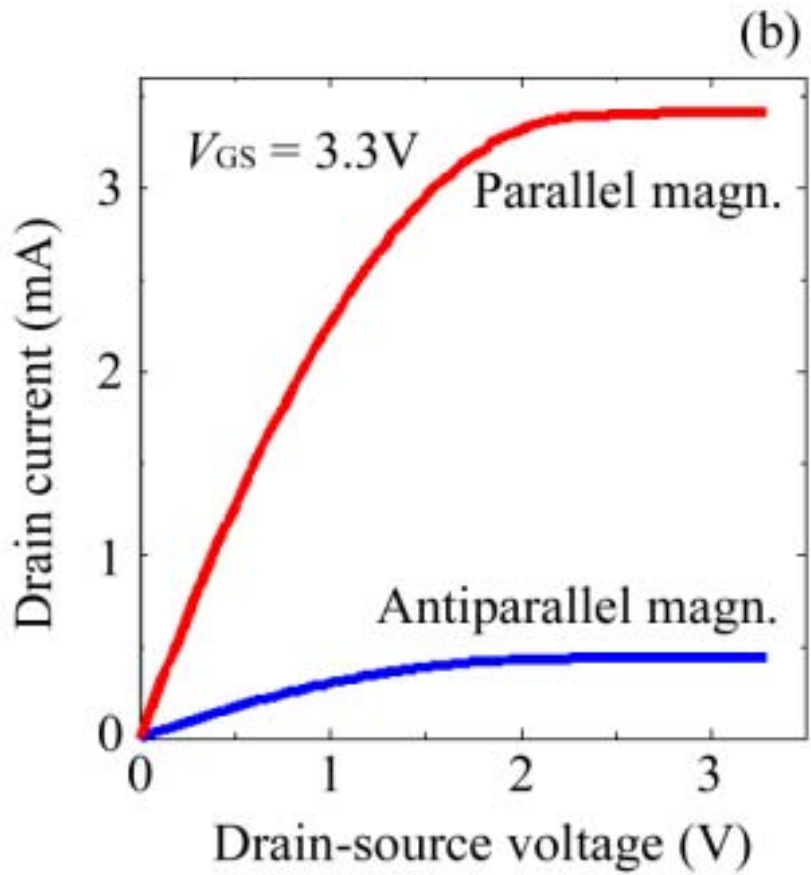

Fig, 1.

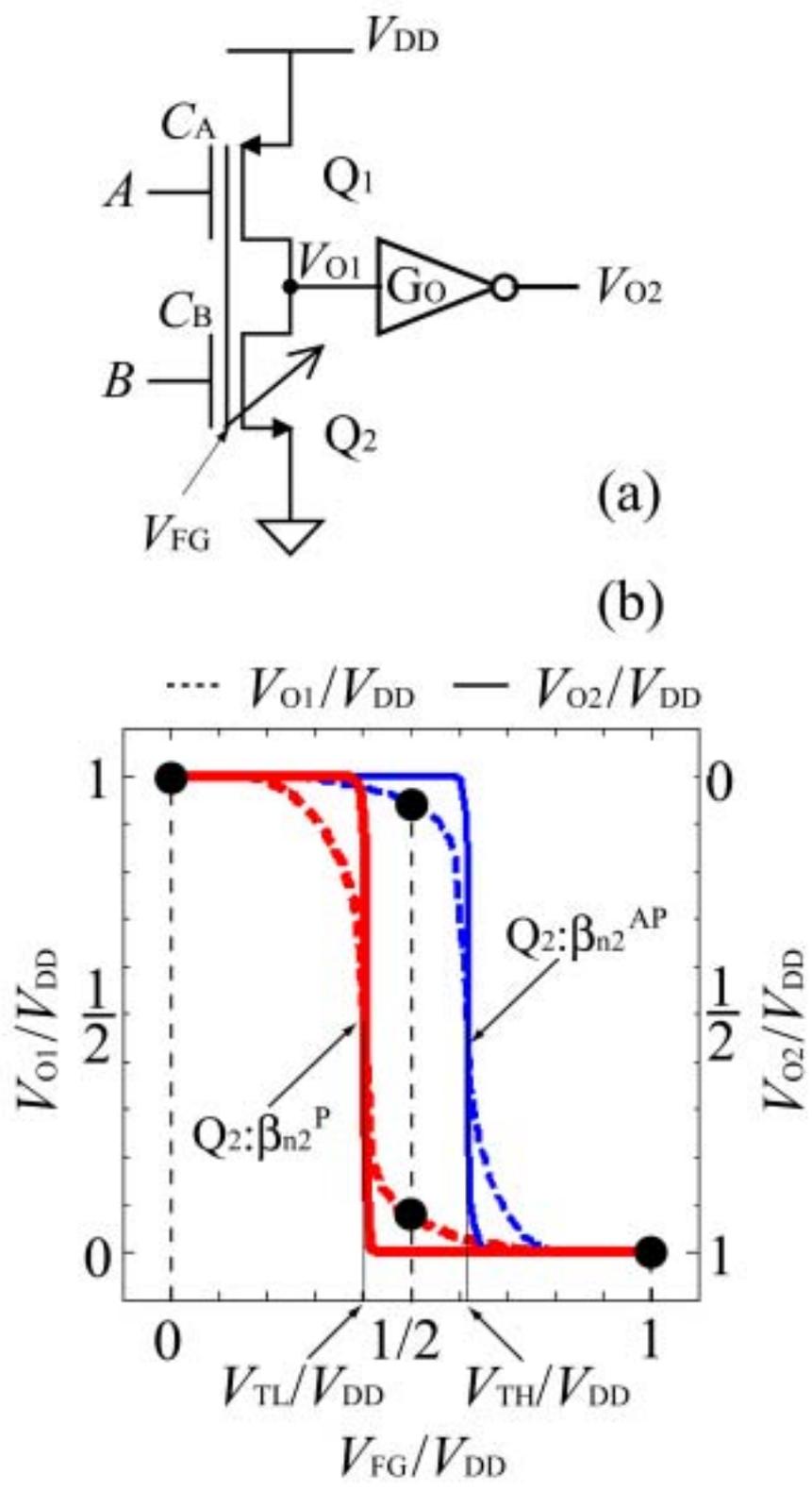

Fig. 2.

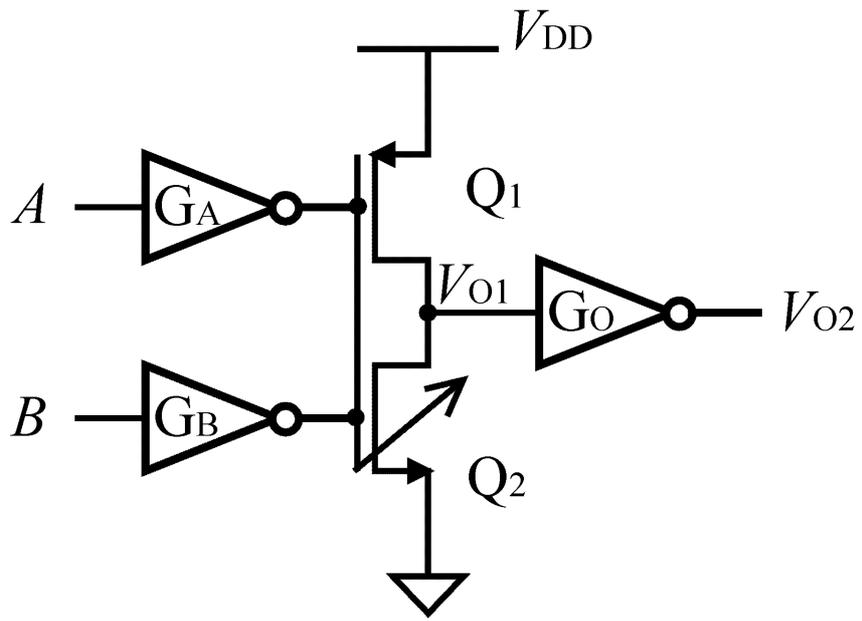

Fig. 3.

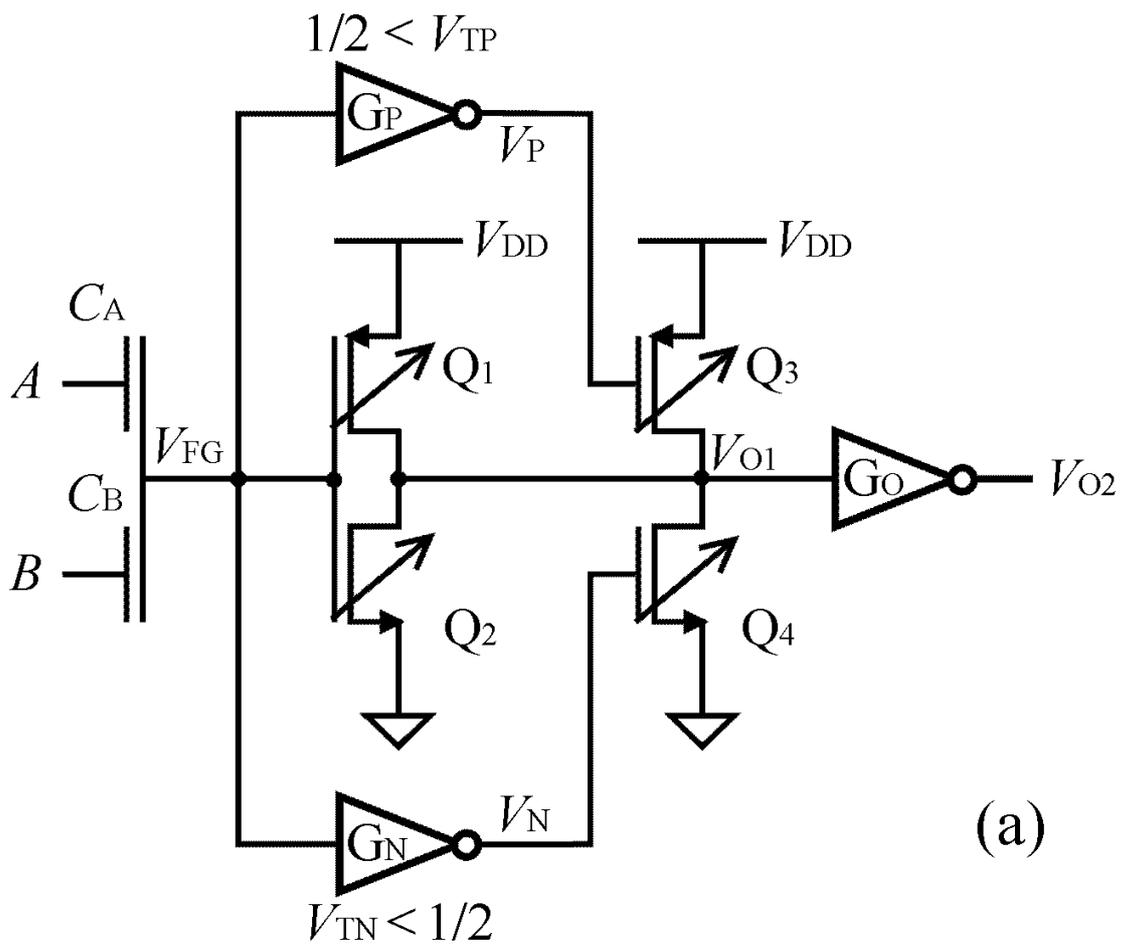

Fig. 4 (a)

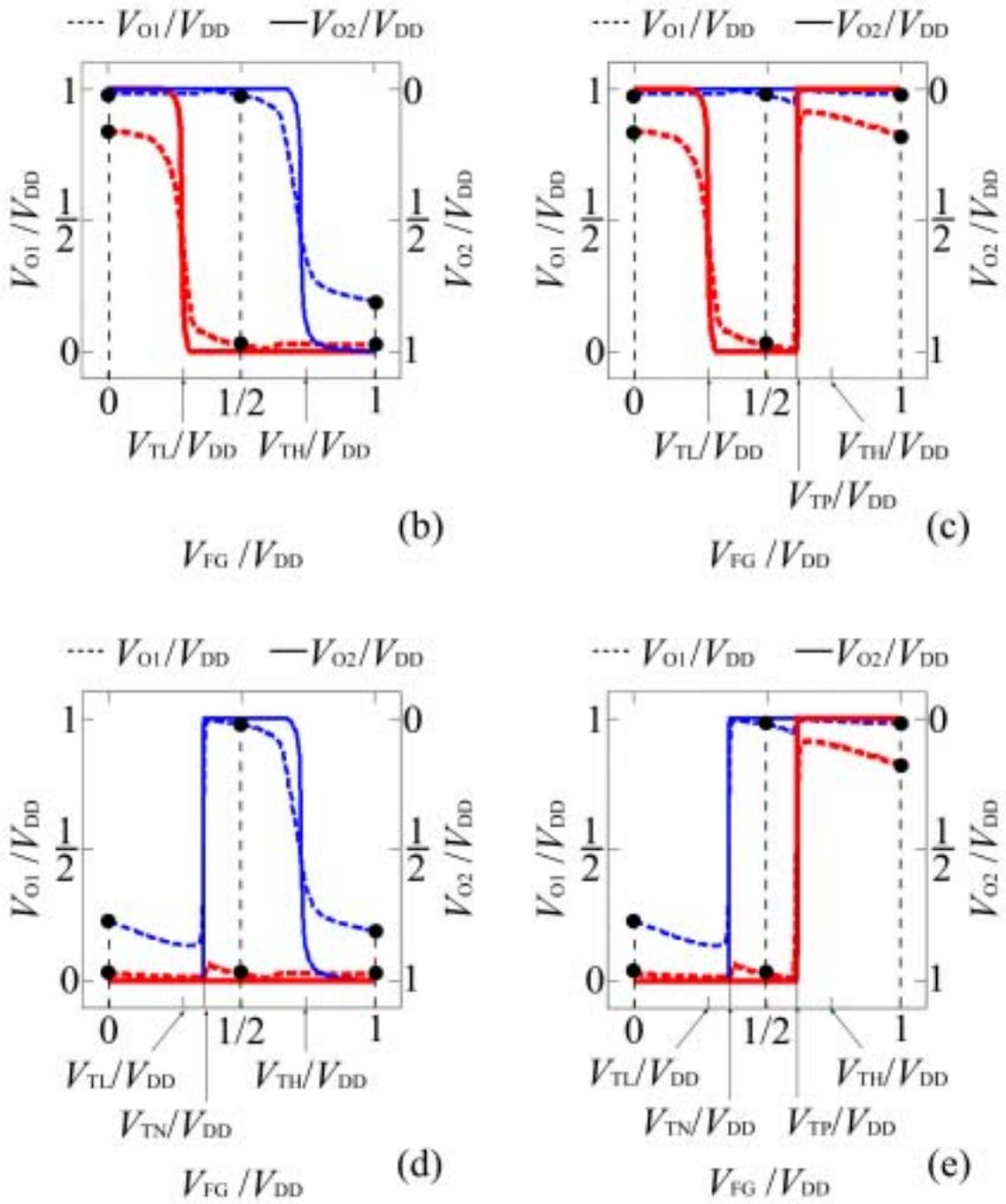

Fig. 4 (b), (c), (d), and (e)

Table I.  Truth table of the reconfigurable circuit of Fig. 4 (a) for each condition.   Tables for (a) $\{Q_1, Q_2, Q_3, Q_4\} = \{\zeta_1, \zeta_2, AP, AP\}$, (b) $\{\zeta_1, \zeta_2, P, AP\}$, (c) $\{\zeta_1, \zeta_2, AP, P\}$ and (d) $\{\zeta_1, \zeta_2, P, P\}$ correspond to Figs. 4 (b), 4(c), 4(d), and 4(e), respectively.

(a)  $\{\zeta_1, \zeta_2, AP, AP\}$

| $\{\zeta_1, \zeta_2\}$ ||| $\{P, AP\}$ || $\{AP, P\}$ ||
|---|---|---|---|---|---|---|
| $A$ | $B$ | $V_{FG}$ | $V_{O1}$ | $V_{O2}$ | $V_{O1}$ | $V_{O2}$ |
| 0 | 0 | 0 | ~1 | 0 | ~1 | 0 |
| 0 | 1 | }1/2 | ~1 | 0 | ~0 | 1 |
| 1 | 0 |  |  |  |  |  |
| 1 | 1 | 1 | ~0 | 1 | ~0 | 1 |

(b)  $\{\zeta_1, \zeta_2, P, AP\}$

| $\{\zeta_1, \zeta_2\}$ ||| $\{P, AP\}$ || $\{AP, P\}$ ||
|---|---|---|---|---|---|---|
| $A$ | $B$ | $V_{FG}$ | $V_{O1}$ | $V_{O2}$ | $V_{O1}$ | $V_{O2}$ |
| 0 | 0 | 0 | ~1 | 0 | ~1 | 0 |
| 0 | 1 | }1/2 | ~1 | 0 | ~0 | 1 |
| 1 | 0 |  |  |  |  |  |
| 1 | 1 | 1 | ~1 | 0 | ~1 | 0 |

(c)  $\{\zeta_1, \zeta_2, AP, P\}$

| $\{\zeta_1, \zeta_2\}$ ||| $\{P, AP\}$ || $\{AP, P\}$ ||
|---|---|---|---|---|---|---|
| $A$ | $B$ | $V_{FG}$ | $V_{O1}$ | $V_{O2}$ | $V_{O1}$ | $V_{O2}$ |
| 0 | 0 | 0 | ~0 | 1 | ~0 | 1 |
| 0 | 1 | }1/2 | ~1 | 0 | ~0 | 1 |
| 1 | 0 |  |  |  |  |  |
| 1 | 1 | 1 | ~0 | 1 | ~0 | 1 |

(d)  $\{\zeta_1, \zeta_2, P, P\}$

| $\{\zeta_1, \zeta_2\}$ ||| $\{P, AP\}$ || $\{AP, P\}$ ||
|---|---|---|---|---|---|---|
| $A$ | $B$ | $V_{FG}$ | $V_{O1}$ | $V_{O2}$ | $V_{O1}$ | $V_{O2}$ |
| 0 | 0 | 0 | ~0 | 1 | ~0 | 1 |
| 0 | 1 | }1/2 | ~1 | 0 | ~0 | 1 |
| 1 | 0 |  |  |  |  |  |
| 1 | 1 | 1 | ~1 | 0 | ~1 | 0 |